\documentclass[letterpaper]{article}
\usepackage{aaai}
\usepackage{times}
\usepackage{helvet}
\usepackage{courier}
\usepackage{graphicx}
\usepackage{amsmath}
\usepackage{amssymb}
\usepackage{setspace}

\usepackage{color}

\newcommand{\revl}[1]{{\textcolor{black}{#1}}}
\newcommand{\revxl}[1]{{\textcolor{black}{#1}}}

\title{Information Diffusion in Computer Science Citation Networks}

\begin{document}

\author{Xiaolin Shi \\  Dept. of EECS \\ University of Michigan \\ Ann Arbor, MI \\  shixl@umich.com
	\And Belle Tseng \\ Yahoo Inc. \\ 3420 Central Expwy \\ Santa Clara, CA \\ belle@yahoo-inc.com
	\And Lada Adamic \\ School of Information \\  University of Michigan \\ Ann Arbor, MI \\ ladamic@umich.edu}

\maketitle
\begin{abstract}
The paper citation  network is a traditional social medium for
the exchange of ideas and knowledge. 
In this paper we view citation networks from the
perspective of information diffusion. We study the structural
features of the information paths through the citation networks of publications in computer science, and analyze
the impact of various citation choices on the subsequent impact of
the article. We find that citing recent papers and papers within the same
scholarly community garners a slightly larger number of
citations on average.

However, this correlation is weaker \revl{among well-cited papers
implying that for high impact work citing within one's field is of lesser importance}. We also study differences in information flow for specific subsets of citation networks: books versus conference and journal articles, different areas of computer science, and  different time periods.

\end{abstract}

\section{Introduction}

Information diffusion is the communication of knowledge over time
among members of a social system. In order to analyze information
diffusion, one needs to study the overall information flow and
individual information cascades in the networks. Although much
recent attention has been focused on new forms of collective content
generation and filtering, such as blogs, wikis, and collaborative
tagging systems, there is a well established social medium for
aggregating and generating knowledge --- published
scholarly work. As researchers innovate, they not only publish new
results, but also cite previous results and related work that their
own innovations are based on. This creates a social ecology of
knowledge --- where information is shared and flows along
co-authorship and citation ties.

In this paper, we examine
information flow \revl{within and between different areas of computer science and its impact}. Our basic assumption is many citations are evidence of \revl{information flow} from one article, and its authors, to another. In order to cite a paper, an author usually, though not always~\cite{simkin2005smc}, reads the paper and acknowledges it as being relevant to the subject of their own paper, either by providing information that their work is built upon, or by providing information about related approaches to the same problem. Although not every citation represents the same level of engagement, citation networks provide some of the clearest evidence of information flow.

Our work has two primary goals: first, we are interested
in observing the features of information flow in citation networks; and second, we
want to know which of these features, such as time \revl{spans} and community
structure \revl{representing different fields of research}, affect the information flow. 

Studying citation networks has been the purview of the field of
scientometrics, which aims to measure the impact of scholarly
publications~\cite{dieks1976dis}. Scientometric data has been available for several decades and so it was already in the 1960s that de Sola Price first
observed power laws in scientific citation networks and developed
models of citation dynamics~\cite{desollaprice1965nsp}.

However, the recent emergence of online knowledge sharing has made it
particularly easy to study information diffusion on a large scale.
Studies of information cascades in blogs~\cite{Kumar03,Adar04,Leskovec07}, social bookmarking sites, and photo sharing have all revealed a highly skewed distribution in the attention a particular post, URL, new story~\cite{lerman2007sip}, or photo~\cite{Lerman07Flickr}  will receive. The attention may be measured through links or tags given to the items. In sepererate studies, it has been shown that such networks
exhibit strong community structure~\cite{tseng2005tcv,adamic2005pba,1149945}, where \revxl{links or interactions} occur more frequently within communities than between them.

The role of community structure in information diffusion has also been studied in scientific citation networks. It has been found that
there is a longer delay for citations across disciplines than ones
within a discipline, implying that information is not only less
likely to diffuse across community boundaries, but when it does, it
will do so with a longer time delay~\cite{rinia2001cdi}. \revl{Information flow between communities is such a relatively small proportion of total information flow, that
modeling citation networks without them} provides realistic citation distributions and
clustering coefficients~\cite{borner2004sea,rosvall07}. The
development of efficient network algorithms has lead not just to discoveries of the
the overall properties of citation networks, but also the detection
of changes in citation patterns where a new trend or paradigm
emerges~\cite{Leicht2007}. There has also been interest in
visualizing and quantifying the amount of information flow between
different areas in science~\cite{boyack2005mbs}, in effect mapping
the generation of human knowledge through information flows. These maps leave open the question, however, of what happens once information has diffused across a community boundary\revxl{; w}ill it have the same impact as information diffusing within a community?

This is an interesting question, because recent empirical work~\cite{guimera2005tam} has shown that new
collaborations between experienced authors are more likely to result
in a publication in a high impact journal than in collaborations
between unseasoned authors or repeat collaborations between the same
two authors. The argument is that merging ideas and expertise in a
novel way will produce higher impact work. But this work did not
address whether the authors were from the same scientific
communities or not, or whether the publications cited in the work
stemmed from the same field.  On the theoretical side, agent based
models of innovation have shown that independent innovation within
communities is important, so that the network as a whole does not
converge on suboptimal solutions too quickly~\cite{lazer2005hat}.

In this paper, to answer the question of the impact of cross-community information flows in computer science, we  make empirical observations of citations of computer science articles, focusing specifically on information flow across community boundaries and temporal gaps. In the following sections, we first describe the computer science publication data sets we used and the construction of the
citation networks. \revl{We then examine the properties of the citation networks, and relate the
properties of a citing link to subsequent impact of the citing article}. 

\section{Preliminaries} \label{sec:preliminaries}

\subsection{Definition of citation networks}

Citation networks are networks of references between documents.
 \revl{In this paper, we focus on
paper citation networks, which  correspond to} information
diffusion in the corresponding research areas.

From the graph theoretic perspective, citation networks can be thought of as directed graphs with time stamps and community labels on each node:
\begin{itemize}
    \item \emph{Nodes}: publications;
    \vspace{-0.1cm}
    \item \emph{Edges}: one paper citing another;
    \vspace{-0.1cm}
    \item \emph{Edge directions}: in order to represent the direction of information flow, we denote the direction of edges from cited papers to citing papers;
    \vspace{-0.1cm}
    \item \emph{Time stamps}: years in which the papers were published;
    \vspace{-0.1cm}
     \item \emph{Time spans}: the time \revl{elapsed between the publication of the cited and citing paper};
     \vspace{-0.1cm}
    \item \emph{Community labels}: we classify the papers     into different research areas according to their venue information.
\end{itemize}

Information flows in citation networks
can be interpreted as the scientific ideas and knowledge transmitted
from publication to publication, which are explicitly indicated by
citation relationships.
\revl{Not all, or perhaps very little, information is preserved from cited to citing paper. Further, the information may be amended in the citing paper. Nevertheless, we assume that the cited paper {\em informed} the citing paper.} There are
two common and significant features of any typical citation network:
first, it is directed and almost acyclic; and second, when it
evolves over time, only new nodes and edges are added, and none
are removed \cite{Leicht2007}. The acyclic nature of the graph stems
from the simple fact \revl{that}, with very few exceptions, a paper will not
cite a paper published in the future. Although publication delays
may lead to such occurrences, most citations are limited to
previously published work.

\subsection{Description of data sets}

The datasets we study are two large digital libraries
encompassing comprehensive scholarly articles \revl{primarily} in computer science --- the \texttt{ACM}\footnote{\texttt{http://portal.acm.org}} data set and the \texttt{CiteSeer} \footnote{\texttt{http://citeseer.ist.psu.edu}} data set \cite{Giles04}.  In the \texttt{ACM} data set, there are several different types of publications, such as books, journal articles, conference proceeding papers, reports, and theses. Books alone account for 113,089 of the publications in the \texttt{ACM} dataset.  Both of the data sets have 
information about the \revl{publication dates and venues}; however,
some of the information is incomplete or inaccurate. Since our study considers the time evolution and community structure of the
networks, we deleted the nodes with an unresolved time or venue
information.

\begin{table*}[tbhp]
\centering
\begin{tabular}[h]{|l||rr||rrr||rrr|}
  \hline
 &  \multicolumn{2}{c|} {Orig.} & \multicolumn{3}{|c|}{With Publication Date}   &  \multicolumn{3}{|c|}{With Publication Venue}  \\
  \hline
  \hline
  & Nodes & Edges & Nodes & Edges & Time range & Nodes & Edges & Communities \\
  \hline
  \texttt{ACM} & 842,422 & 2,492,503 & 250,556 & 861,088 & 1920 - 2005 & 119,268 & 346,289 & 26 \\
  \hline
  \texttt{CiteSeer} & 716,774 & 1,438,505 & 93,298  & 342,657 & 1958 - 2005 & 52,411 & 84,134 & 23 \\
  \hline
\end{tabular}
\caption{Summary statistics of the citation networks before and
after cleaning.} \label{table:summary}
\end{table*}

\revl{While \texttt{ACM} data set includes citations to
publications outside of the data set, the \texttt{Citeseer} data does not, and so we limit our analysis to citations between articles within each dataset. In addition, some citations between two articles that both reside in the same data set are missing, due to the difficulty in disambiguating and parsing citations from article text~\cite{simkin2005smc}.  Even with these limitations, we are left with 346,000 citations for the \texttt{ACM} dataset and 84,000 citations in the \texttt{CiteSeer} dataset, which we use to measure information flows between different computer science communities and the impact of a publication. We will discuss possible biases introduced by missing data below.}

Even though we are analyzing two separate datasets, they overlap in subject area and time span. It is therefore reassuring that they have a significant, but relatively small overlap in the articles that they contain. There are 613,444 proceedings or journal papers in the \texttt{ACM} dataset that we are studying, and 593,386 of them have distinct titles in the database; while there are 716,774 papers in \texttt{CiteSeer} dataset, and 611,127 have distinct titles. By matching the titles and authors of the 593,386 papers in \texttt{ACM} and 611,127 papers in \texttt{CiteSeer} using a simple cosine similarity measure, we identify 122,978 (20\%) papers that are present in both datasets. Finally, Table \ref{table:summary} gives summary statistics of the two data sets and the citation networks we will study.

\section{Structural features of citation networks}\label{sec:descriptive}

Since the structural features of citation networks provide explicit
evidence of information flow paths, our study of
information diffusion starts with them.

\subsection{Degree distributions}\label{sec:degdist}

As stated before, we \revl{set the direction of an edge to reflect the
direction} of information flow. The in-degree is the number of
papers cited in a given paper. In effect, it is the number of papers
that may have influenced the paper at hand. The out-degree is the
number of papers citing the given paper, reflecting the paper's potential
impact and influence.

\begin{figure}[h]
  \centering
\begin{tabular*}{\columnwidth}{@{\extracolsep{\fill}}cc}
  \includegraphics[width=0.5\columnwidth]{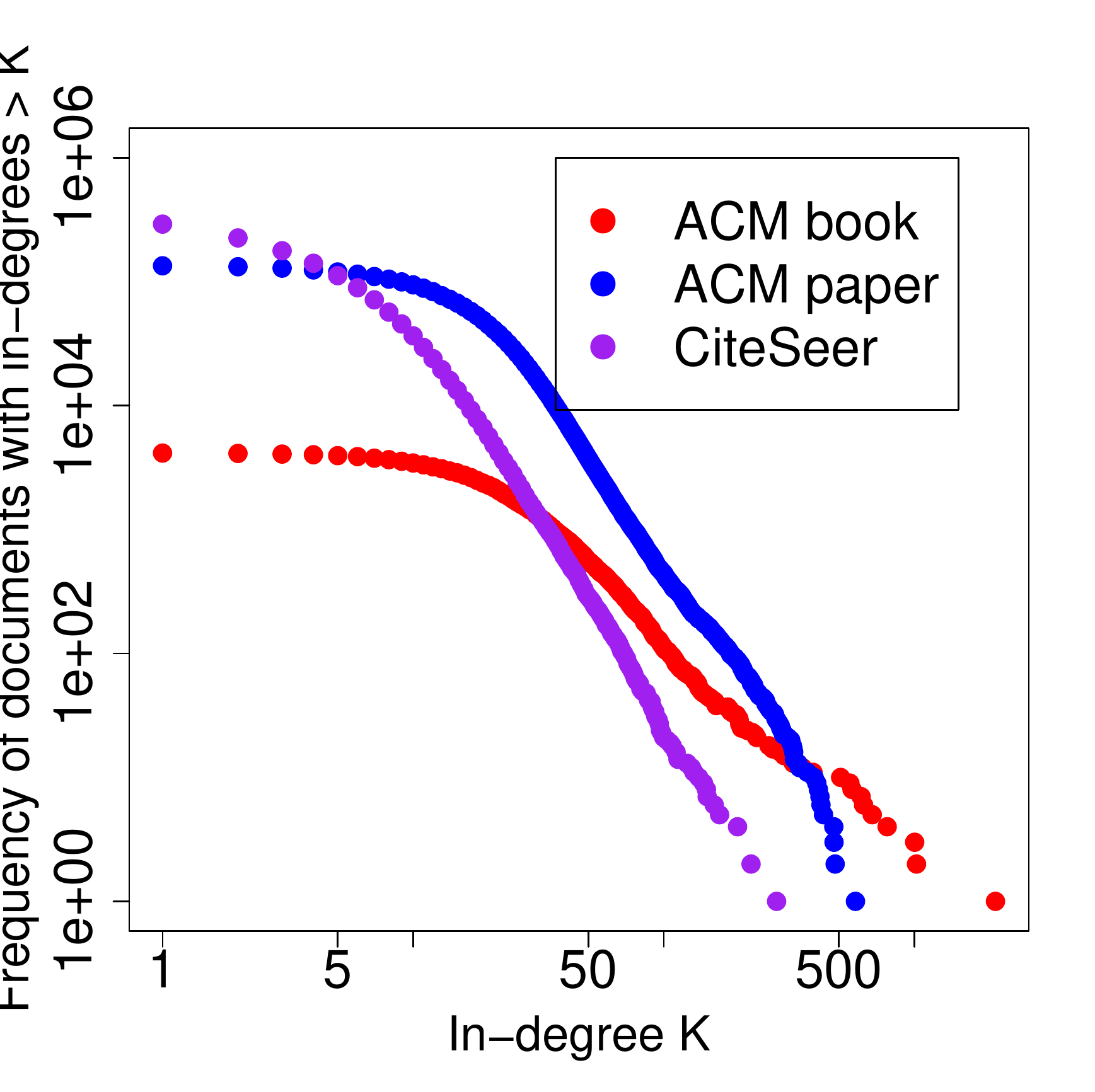} &
\includegraphics[width=0.5\columnwidth]{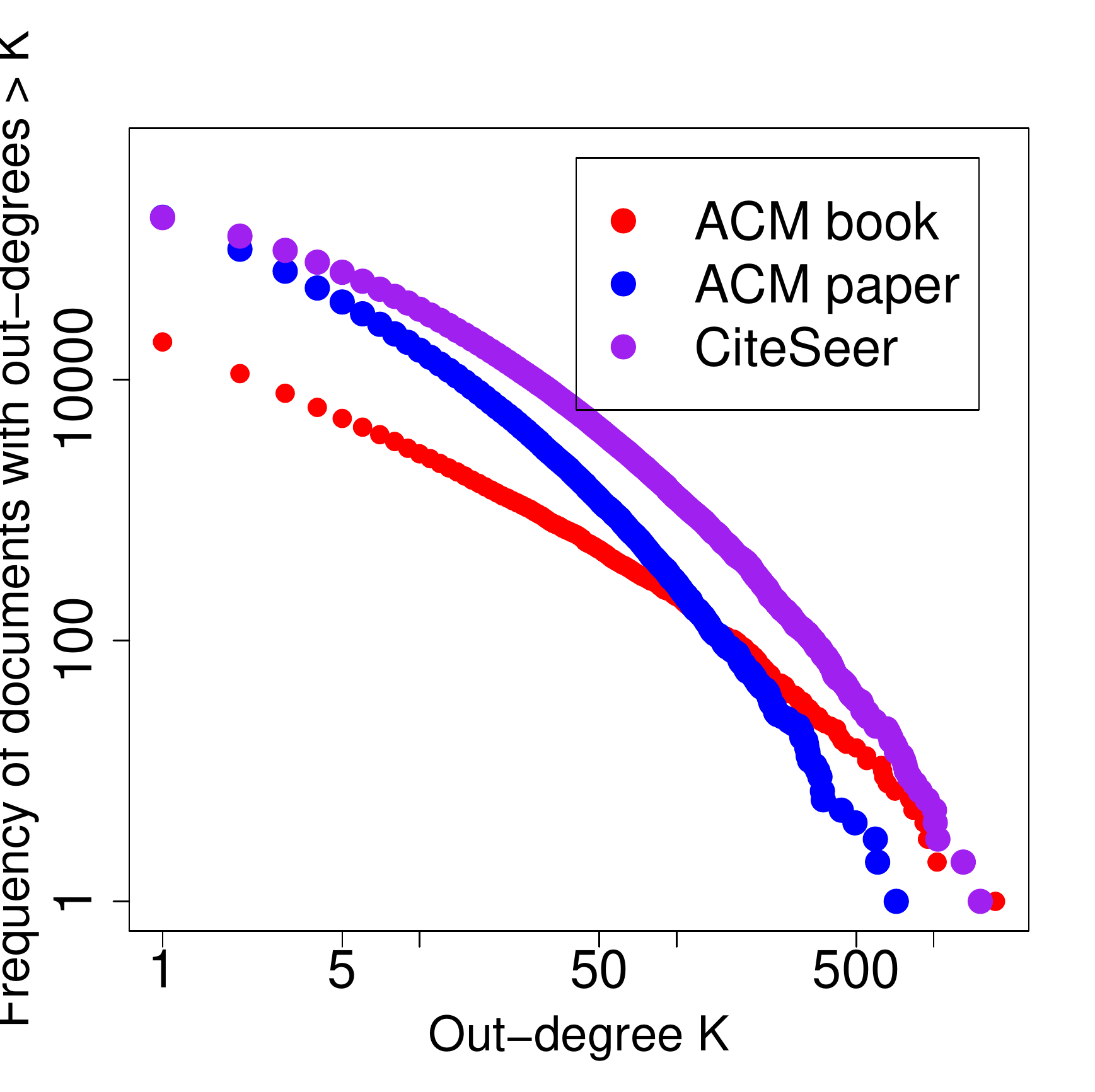} \\
 (a) In-degree distributions &  (b) Out-degree distributions \\
\end{tabular*}
  \caption{The degree distributions of the \texttt{ACM} and \texttt{CiteSeer} citation graphs.}
  \label{fig:degreedistributions}
  \vspace{-0.3cm}
\end{figure}

In the previous section, we mentioned that there are different types of publications in \texttt{ACM}, including those with very high in-degrees, such as books. Since these comprehensive publications normally have many more references than regular papers, we show the degree distributions separately for books and papers in the \texttt{ACM} data set in Figure \ref{fig:degreedistributions}. \revl{The distribution of the length of the reference list for publications (their indegree) is  highly skewed;} some publications have references from 10 to several hundred papers in the dataset, many have none, or few. In actuality, many of these papers have longer lists of references, but these were not identified, or they fall outside of the dataset. The distributions of out-degrees \revl{are similarly skewed, an indication of a linear preferential attachment mechanism}: already well cited papers are more easily discovered, and subsequently cited: it is the success-breeds-success phenomenon \cite{Burrell03}.  \revl{As one might expect, both the in-degree distributions and out-degree distributions of books in the} \texttt{ACM} dataset are significantly heavier tailed -- with books both citing more and being cited more. It is therefore unsurprising that the in-degree and out-degree distributions of documents in \texttt{CiteSeer} are more similar to those of \texttt{ACM} papers, as opposed to books.

\subsection{Connectivity}

In order to analyze how information flows \revl{through} the citation
networks, we first study their connectivity.

Two vertices $A$ and $B$ are said to be in the same \emph{strongly connected component} if there exists a path both from $A$ to $B$ and from $B$ to $A$. For both the \texttt{ACM} and \texttt{CiteSeer} citation graphs with
correct time stamps, there are no significant strongly connected components. The absence of large strongly connected components is consistent with the fact that citation graphs are nearly perfect directed acyclic graphs (DAGs).
However, $96.08\%$ of the nodes (publications) 
are in the largest \emph{weakly connected component}, in which there is a path between every pair of nodes in the version of undirected graph, in the \texttt{ACM} citation network. Similarly, $96.20\%$ of the nodes 
are in the largest weakly connected component of the \texttt{CiteSeer} citation network.

In the \texttt{ACM} dataset,
5.0\% of the papers are published in the years 2004 and
2005. By tracing back their citations, we find that
48.8\% of the papers published in previous years are either
directly or indirectly cited by the papers in 2004 and 2005. In
the \texttt{CiteSeer} data set, with  
3.5\% papers published in
the years 2003 - 2005, 28.9\% of earlier papers are reachable by
tracing back the citations. 
Although the two data sets differ in their cohesion (either due to completeness of data, or other issues of coverage), we observe \revl{that} each subsequent generation of papers is tied directly or indirectly to a significant portion of the prior work.

\begin{figure}[htbp]
  \centering
  \begin{tabular*}{\columnwidth}{@{\extracolsep{\fill}}cl}
  \includegraphics[width=0.5\columnwidth]{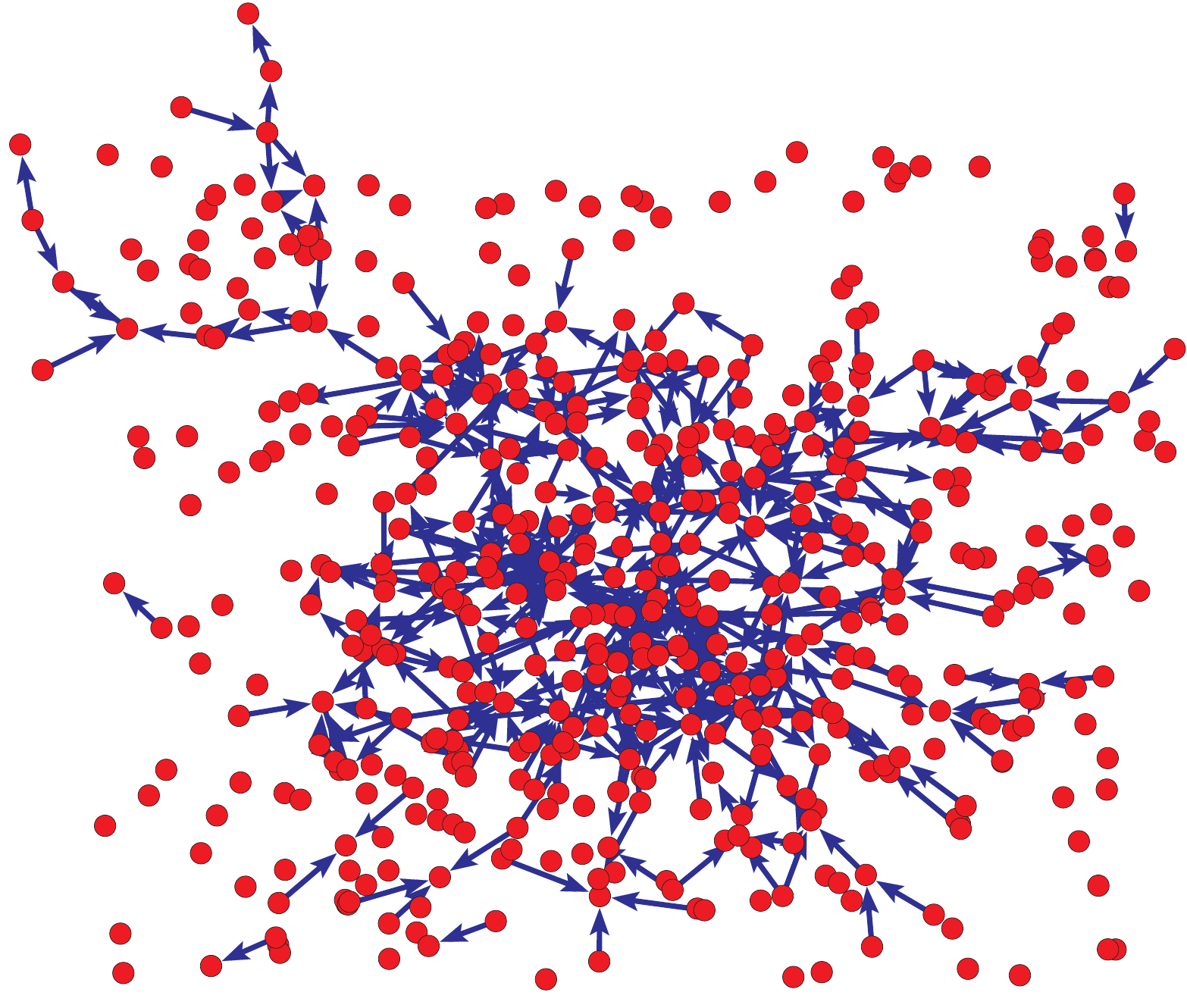}  &
    \includegraphics[width=0.48\columnwidth]{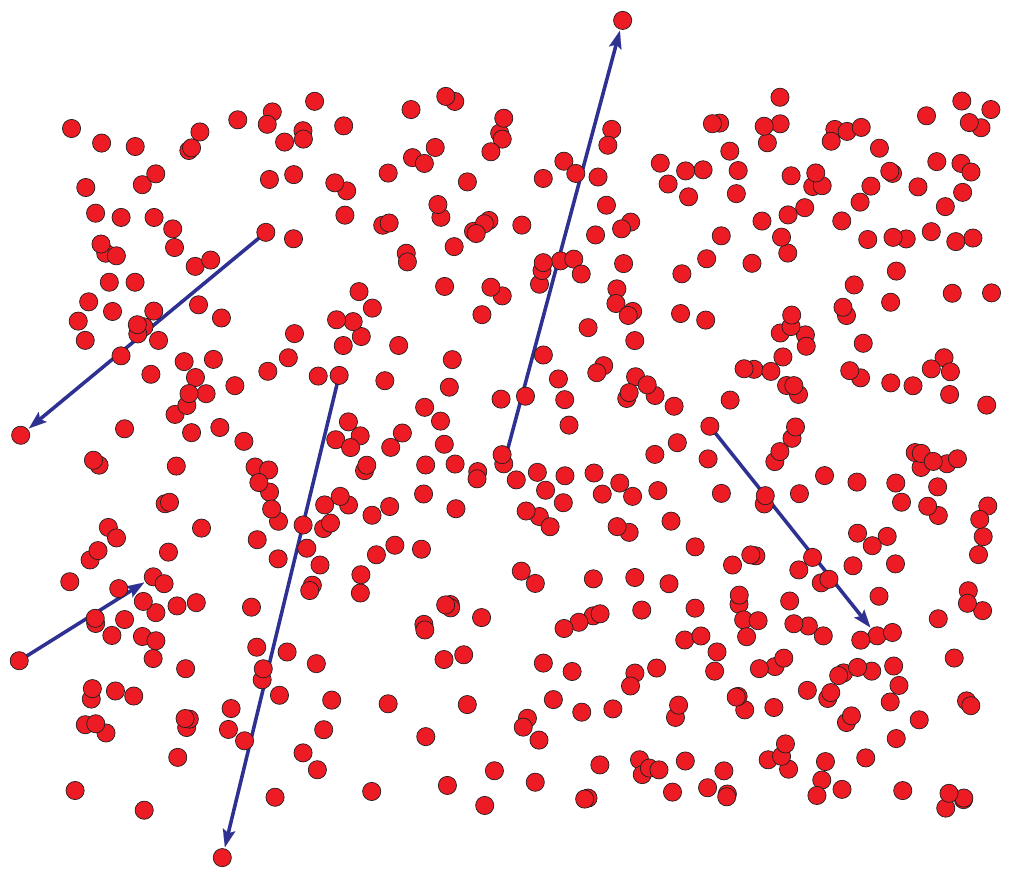} \\
    (a) 500 most cited papers & (b) 500 random papers
  \end{tabular*}
  \caption{Subgraphs of the 500 most cited papers and 500 random papers in the \texttt{ACM} citation network.}
  \label{fig:ACM500papers}
  \vspace{-0.1cm}
\end{figure}

If we take the 500 most cited articles in the \texttt{ACM} network, we observe that there is a giant component linking a significant fraction of the most
influential papers (see Figure \ref{fig:ACM500papers} (a)). In
contrast, if we were to select 500 random papers, there would be no
giant component - as papers are rather unlikely to cite one another
(Figure \ref{fig:ACM500papers} (b)).  This observation is
consistent with many other networks where the most highly connected
nodes tend to be connected to one another \cite{Shi2008}. It is \revl{also} known as the rich-club phenomenon \cite{Zhou04,Colizza2006}. Yet it is still striking that such a small number of most influential papers out of tens of thousands in computer science should be connected \revl{to one another through one another}.

\subsection{Average shortest directed path}

The shortest paths in graphs (also termed geodesics) relate directly to the accessibility of information. Like many other complex
networks, the citation networks we study exhibit the
\emph{small world phenomenon}. The average shortest directed path of the
\texttt{ACM} graph is 7.60, and its largest geodesic is 32. Similar
to the \texttt{ACM} citation network, \texttt{CiteSeer} has an
average shortest directed path of 6.29 and its longest geodesic is
28. However, the  reachable pairs of nodes via directed paths in the
\texttt{ACM} citation network \revl{comprise} $0.65\%$ of all possible paths (or node pairs) and
$0.41\%$ of all possible paths in the \texttt{CiteSeer} citation
network.

From the connectivity and shortest directed paths of the citation
networks, we can see that, in spite of the largest weakly connected
component occupying nearly the entirety of the network, the percentage of reachable pairs of nodes is smaller. But where paths do
exist, we observe that the lengths of information flows are generally short. Note that this does not preclude that there are more circuitous routes involving several papers. In our measurements, we only account for the shortest path. This form of ``deep linking", citing original articles, considerably shortens the path between articles.

\subsection{Sizes of information cascades \label{sec:cascade}}

\begin{figure}[htbp]
  \centering
  \includegraphics[width=0.7\columnwidth]{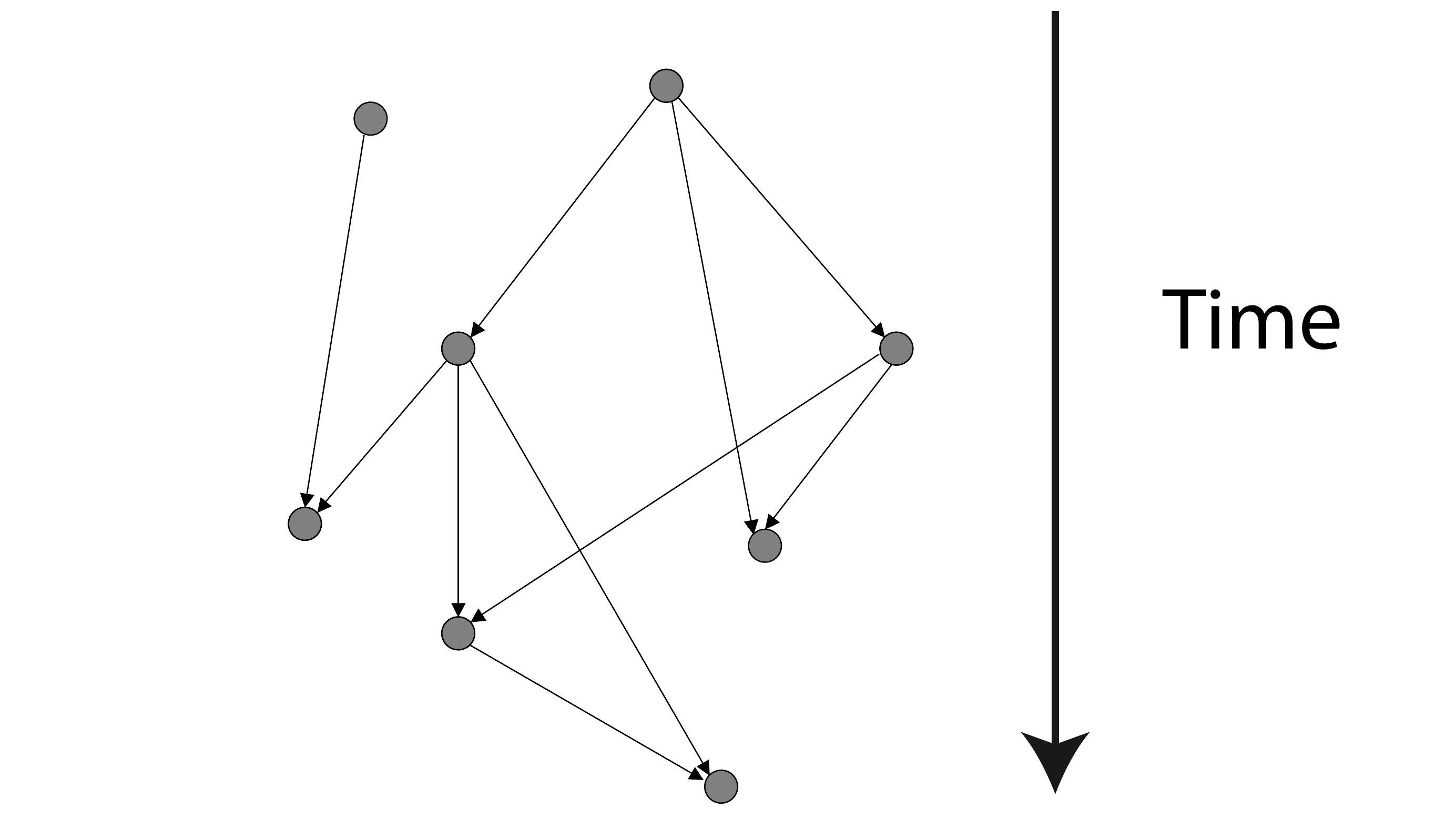}
  \caption{Illustration of a information cascade with multiple temporal levels.}
  \label{fig:cascade}
  \vspace{-0.1cm}
\end{figure}

One may be interested not only in direct citations of a given
paper, but subsequent citations of the citing papers, etc. Figure \ref{fig:cascade} gives a
simple example of an information cascade. An
information cascade represents both the direct and indirect
influence of a given publication, although the influence is diluted
with each subsequent step. As each paper cites several others, it is
difficult to attribute influence to any given chain of citations. It
may be that the reason that A cites B is unrelated to  the reason
why B cited C.

\begin{figure}[htbp]
  \begin{tabular*}{\columnwidth}{@{\extracolsep{\fill}}cc}
  \includegraphics[width=0.5\columnwidth]{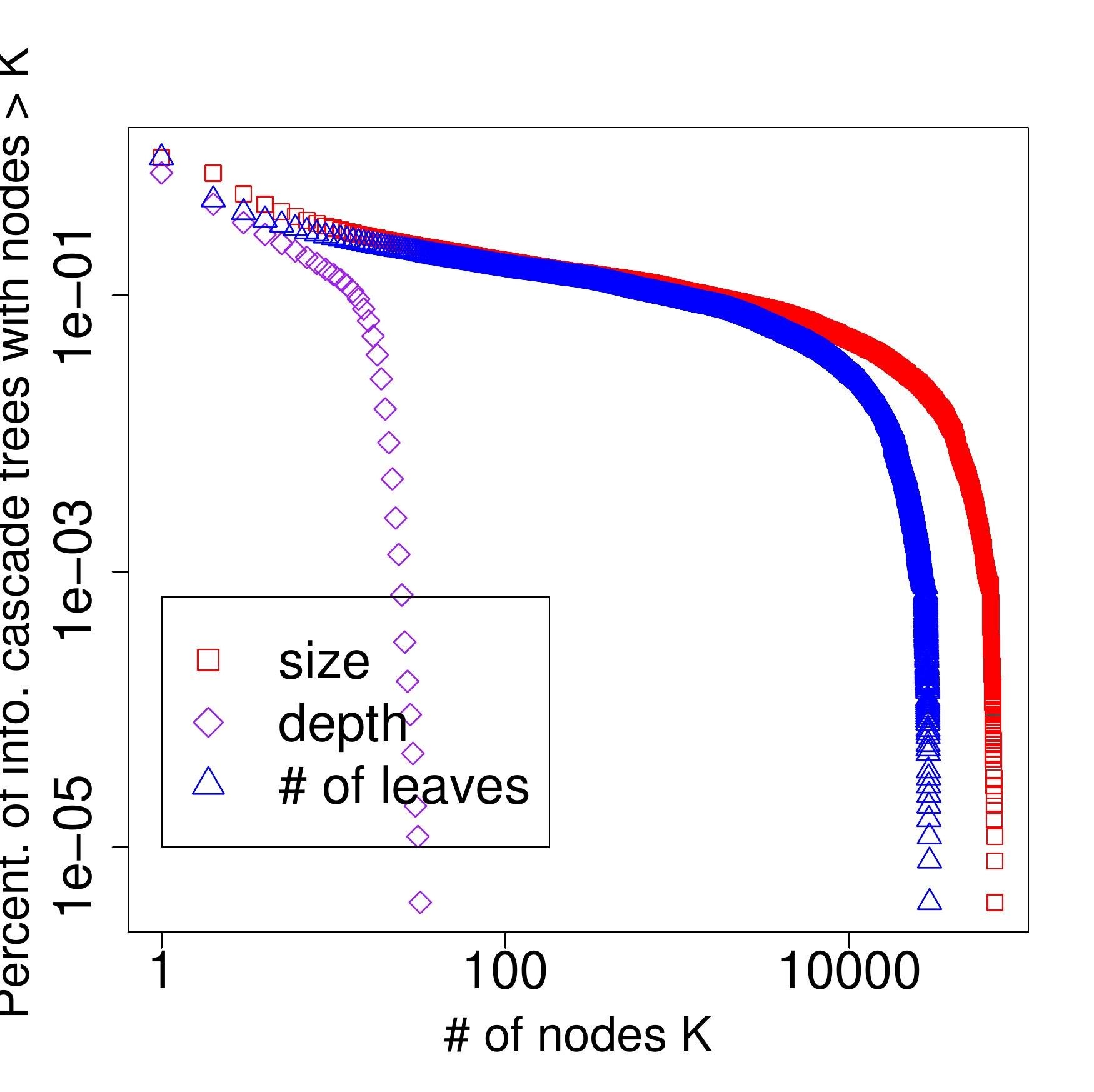} &
  \includegraphics[width=0.5\columnwidth]{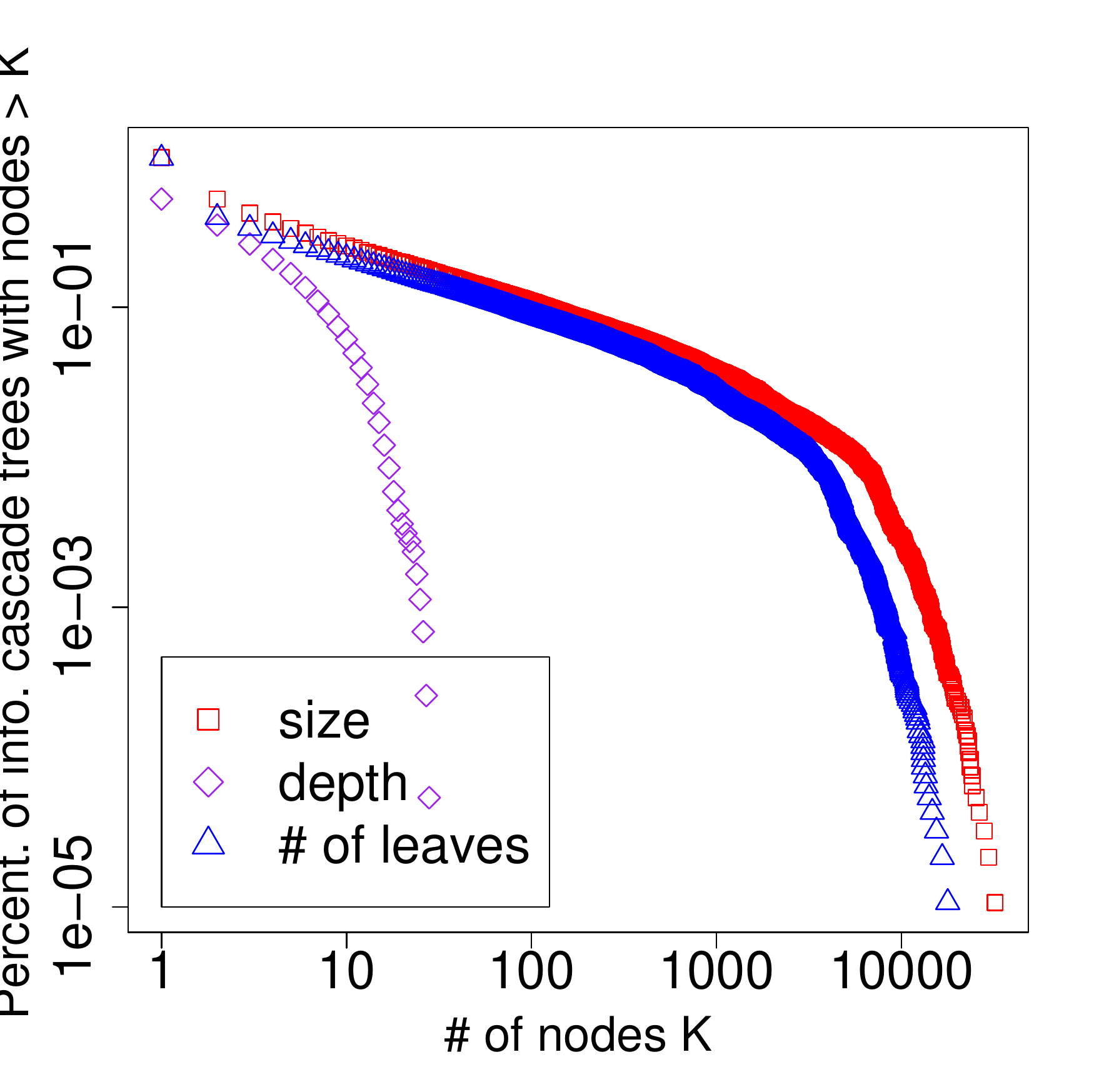} \\
   (a) ACM & (b) CiteSeer \\
   \end{tabular*}
   \caption{The distributions of sizes, depths and numbers of leaves of the information cascade trees in \texttt{ACM} and \texttt{CiteSeer} citation graphs.}
   \label{fig:informationcascades}
   \vspace{-0.1cm}
\end{figure}

By taking every node in the citation graph as a root, we run a breadth first search along the out-going edges, and obtain an information cascade tree starting from every paper in the network. The distributions of sizes, depths and numbers of
leaves of the information cascade trees are shown in Figure
\ref{fig:informationcascades}.

From the figure, we see that all of the distributions of
sizes, depths and numbers of leaves have \revl{a} very a sharp drop in the
tail. These drop-offs naturally correspond to the limits of the data
sets: there are no more than several hundreds of thousands of papers in each
data set, and a cascade can only encompass papers appearing after
the given paper. This is different from the power-law distributions
of the cascade sizes in the blogosphere observed by Leskovec at
al.\cite{Leskovec07}, with slightly different cascade definitions.
The blog measurements were unaffected by size limitations in the
data, as the largest cascade sizes encompassed no more than a few
thousand posts out of millions that were observed.

The Spearman correlations of the cascade sizes, depths and numbers
of leaves in the information cascades, as well as the out-degrees
are shown as Table \ref{table:cascadecorrelations}. We see that, \revl {in both} the \texttt{ACM} and \texttt{CiteSeer} citation graphs, the sizes
and depths of the information cascade trees have large
correlations, meaning that cascades that encompass several generations of scholarly work are also the largest.

\begin{table}[h]
\centering
\begin{tabular}[h]{|l||llll|}
  \hline
  & size \& & size \& & size \& & depth \& \\
  & out-deg & depth & \# leaves & \# leaves \\
  \hline
  \hline
  ACM & $0.724^{***}$ & $0.928^{***}$ & $0.885^{***}$ & $0.802^{***}$ \\
  \hline
  CiteSeer & $0.809^{***}$ & $0.952^{***}$ & $0.884^{***}$ & $0.831^{***}$ \\
  \hline
\end{tabular}
\caption{Spearman correlations of the sizes, depths, numbers of
leaves of information cascades and out-degrees of all papers in the two citation networks. $^{***}$, $^{**}$, and $^{*}$ denote significance at the $< 0.05$, $ < 0.01$ and $\geq 0.01$ levels respectively.}
\label{table:cascadecorrelations}
\end{table}

\section{Information diffusion and the effects of citations}\label{sec:results}

After examining the structural features and the information flow
paths between papers in the citation networks, we turn to how information flows
between communities, and how different types of citations (from and to
various communities and citing old or new papers) would affect the subsequent information diffusion in citation networks.

\subsection{Information flows between communities}\label{sec:infocommunity}

\begin{figure*}[tb]
  \includegraphics[width=0.99\textwidth]{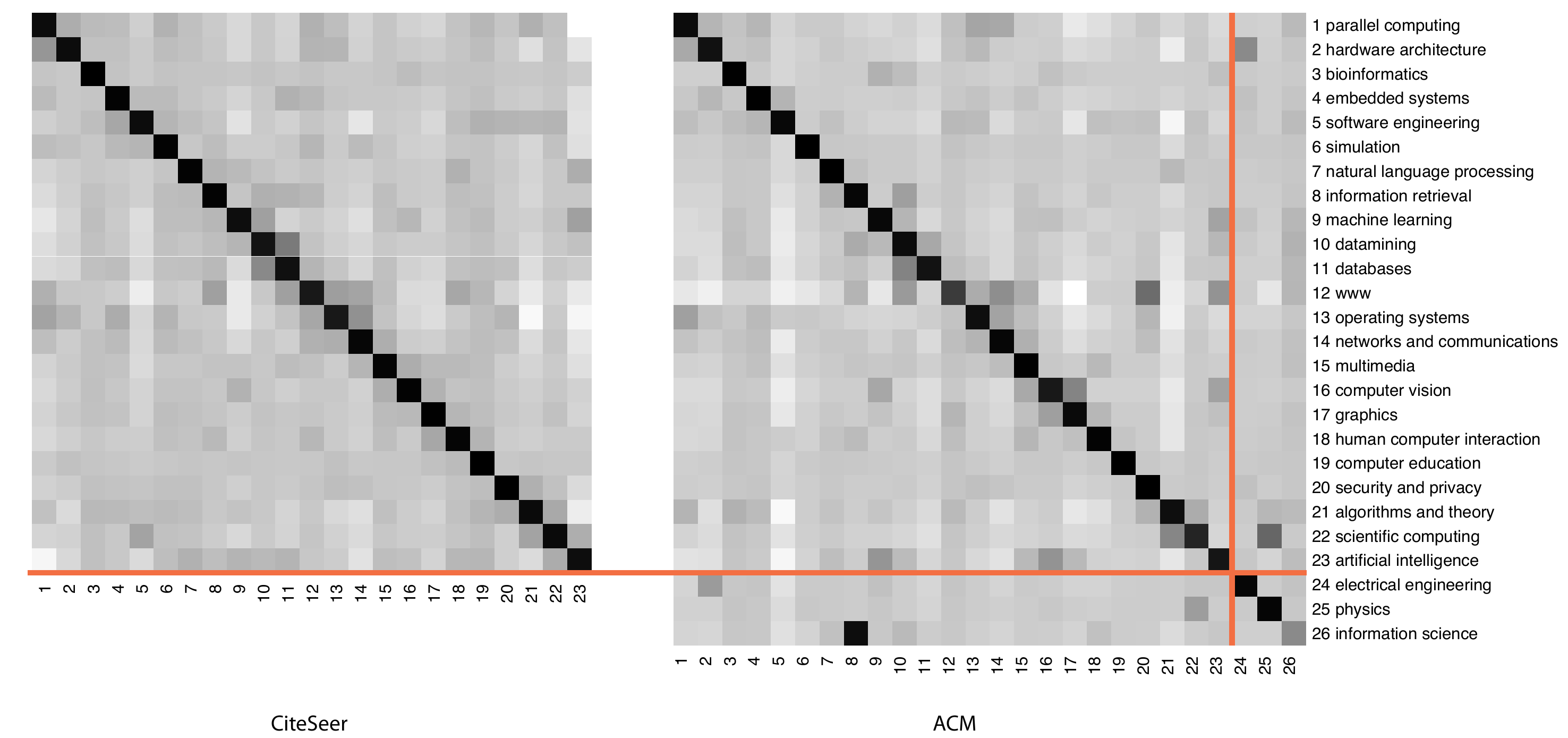} \\
  \vspace{-0.4cm}
   \caption{Visualization of the matrices of community weights between different areas of computer science. Darker cells represent more frequent citation than expected if citation were at random, lighter ones depict less frequent citation.}
   \label{fig:communitymatrices}
\end{figure*}

We assign papers to communities according to their venues, using the classification system adopted by Microsoft's,
\emph{Libra} academic search service\footnote{\texttt{http://libra.msra.cn}} . For example, a paper published in the KDD (\emph{Knowledge Discovery and Data Mining}) Conference would be classified under ``Data Mining", while a paper published in the \emph{Journal of Information Professing and Management} would be classified under ``Information Retrieval".
Because of the incomplete and noisy information in the
venues, we are able to classify about $1/3$ of the papers with about
$80\%-90\%$ precision. \revxl{With this community classification, there are about 205,000 within community citations and 141,000 across community citations in \texttt{ACM}, while 42,000 both within and across community citations in \texttt{CiteSeer}.}

In order to quantify the densities of information flow from community
to community, we first count the number of citations between every
pair of communities for each data set separately (e.g. the number of
citations of Theory to Theory, Theory to Data Mining, etc.), and get a matrix $A$ with these numbers as its entries. \revl{We then} compare the number of citations between any pair of communities relative to the rate of citation we would expect if the volume of inbound and outbound citations were the same, but the citations were allocated at random. We let $N_{ij}$ be the actual number of citations from $i$ to $j$,  $N_{i\centerdot} = \sum_j N_{ij}$ be the total number of citations from community $i$, $N_{\centerdot j} = \sum_i N_{ij}$ be the total number of citations to community $j$, and $N = \sum_{ij} N_{ij}$ be the total number of citations in matrix $A$. Then the expected number of citations, assuming indifference to one's own field and others, from community $i$ to community $j$ is $E[N_{ij}] = N_{i\centerdot} \times N_{\centerdot j}/N$.  We define the community weight as a z-score that tells us how many standard deviations above or below expected $N_{ij}$ is. Here we have the observation that $N \gg N_{i \centerdot}$ and $N \gg N_{\centerdot j}$, so we  approximate the standard deviation by $\sqrt{E[N_{ij}]}$. In this way, for every entry, we get a normalized value, which \revl{we call} \emph{community weight}:
$$
W_{ij} = ( N_{ij} - \frac{N_{i \centerdot} \times N_{\centerdot j}}{N}) /\sqrt{\frac{N_{i \centerdot} \times N_{\centerdot j}}{N}}
$$

By visualizing the normalized matrix, i.e. matrix of community weights, as in Figure \ref{fig:communitymatrices}, we can observe different densities of information flow amongst communities. For example, for each community, as expected, the majority of citations are within the community itself. However, there are some
closely related communities. For example, there appears to be
considerable information flow from Information Science to
Information Retrieval, from Databases to Data Mining,
from Information Retrieval \revl{to} Data Mining and from Computer Vision to
Computer Graphics. These flows reflect
frequent citations by papers from the second community to those in
the first. We also observe that the more theoretical areas such as
Algorithms \& Theory and Physics \revl{are} less connected with
others, while more applied areas, such as Data Mining,
Information Retrieval, and Operating Systems have more
information flows two and from other areas. 

\subsection{Correlations of information diffusion and citation features}

If we define information diffusion to occur when a paper is cited,
then many factors affect such information diffusion. They include
the popularity of the research field pertaining to the article in a
certain period, the reputation of the authors, the specific
innovation reported in the publication, etc. However, there is much
we can surmise simply from the citation patterns, time lapses and
community information. Specifically, we examine what kinds of
citations would make the citing papers have greater impact, whether it is citing another paper in a related community
with strong information flow, or the time elapsed since the publication
of the cited paper.

As we have stated before, to measure the influence of a particular paper, both directly and indirectly influenced papers
may need to be taken into consideration, possibly weighing them
differently. However, for both the clarity of the model and \revl{lack of consensus in the literature for a particular weighting scheme~\cite{aksnes2006cra}}, we use the number of citations a paper receives normalized by the average number of citations received by all papers in the same area and year~\cite{valderas2007}. This measure allows us to make a fair comparison between articles that may not have finished accumulating citations due to their recency, and to account for differences in the publication cycle for different areas~\cite{stringer08}.

\subsubsection{Citation networks for all of computer science}

Since our study focuses mainly on the relationship between information flow and innovation, as opposed to summaries and reviews, we exclude publications that are book chapters and books, and focus on journal articles and papers published in conference proceedings. In the \texttt{ACM} dataset, the articles are already classified according to publication venue type, and so are easily filtered. In the \texttt{CiteSeer} dataset, we find that a majority of publications having 40 or more references tend to be review \revl{manuscripts}. We exclude such publications from both data sets.  Finally, we exclude papers published after 2000, because their recency means that they have not accumulated most of their citations \cite{stringer08,Burrell03}. 

\begin{table*}[htb]
\centering
\begin{tabular}[h]{|l||lll|lll|}
    \hline
   & ACM &  & &   CiteSeer & &  \\
  \hline
  \hline
  & Overall & $\leq$ 90\% & $>$90\%& Overall & $\leq$ 90\% & $>$90\%	 \\
  \hline
  time-diff & $-0.0659^{***}$ & $-0.0581^{***}$ & $0.0045^*$ & $-0.0870^{***}$ & $-0.0899^{***}$  & $0.0124^*$  \\
 \hline
  c-weight & $0.0889^{***}$ & $0.0832^{***}$ & $0.0089^{*}$  & $0.0622^{***}$ & $0.0621^{***}$ & $0.0314^*$ \\
  \hline
  \end{tabular}
\caption{Spearman correlations show the effects of  \revxl{community}
weights and time differences \revxl{between the cited and citing papers} on the \revxl{subsequent impacts of citing papers.}} \label{table:correlations}
\vspace{-0.4cm}
\end{table*}

Table~\ref{table:correlations} shows the correlations between community weights and time lapse of the citing and cited paper, and the subsequent impact of the citing paper. From the table we see that for both   citation networks, the weights of information flows between communities (i.e. the community weights) have positive correlations with the influence metric (normalized out-degrees). This means that, on average, a computer science paper will be rewarded for referencing other papers within its own community or proximate communities. 

\revl{More recent papers have had an opportunity to cite more distant papers in time. Since pairs of citations are only recorded between papers in the dataset, older papers will have shorter recorded timelags to the papers they reference, since earlier referenced papers may not be included. The above is reflected in the correlation between the publication year of the citing paper and the time elapsed between the two papers ($\rho= 0.2, p < 10^{-16}$). More interestingly, there is a negative between the time elapsed between the papers and the subsequent impact of the citing paper. Note that we are already normalizing by the average citation number of papers in a given year, so that older papers' chance to accumulate more citations is not a factor. The negative correlation between citation time lag and impact could be interpreted as citing more recent work  being rewarded by citations}.

However, it is not uncommon to see some extremely innovative and
influential work whose citations reach across communities, or draw upon
older publications. The overall correlations only reflect the
average trend.  \revl{As we observed in Figure~\ref{fig:degreedistributions}, a} large proportion of the papers receives very few citations, while a few papers garner large numbers of them. We found interesting trends, when, in addition to measuring the overall correlation for all papers, we computed separate correlations for the bottom 90\% of the papers according to impact (denoted as $\leq 90\%$ in Table~\ref{table:correlations}) and the top 10\% ($> 90\%$ ).

What we can observe is that for less well cited papers, the
correlations between impact and community information flow weight are positive, in
agreement with the overall trend. This is where the majority of
papers lie --- they receive few citations and do not lead to large
subsequent impact. However, for papers with high impact (dozens
to hundreds of citations), the neutral correlations show that citing within one's own community is less important.

Similar patterns are observed for time lags as well. The lower
impact articles benefit from citing recent work; but for more influential papers, these correlations are reduced or absent. It may be that a truly innovative
article draws upon work that had not been garnering much attention
recently, and that is not tied to many other relevant publications.
This would imply that the more innovative and more
highly cited papers may cross boundaries where information normally
does not flow.

\subsubsection{Subnetworks of papers in different areas}

We have seen how the weights of information flows between communities affect the subsequent impact of the citing papers in the overall citation networks of computer science. In this subsection, we investigate the correlations in a finer scope. We choose the areas whose papers constitute more than 5\% of the total number of papers with community information in both data sets, such as Theory, Distributed and Parallel Computing, Software Engineering, etc. \revl{We consider these papers and their sets of references}. Again, we compute the correlations of the community weights of those citation edges and the normalized out-degrees of these citing papers. The correlations are given in Table~\ref{table:communitycorrelations}. The results show that  the correlations are mostly positive or neutral in three areas, except for papers in theory and algorithms in both \texttt{ACM} and \texttt{CiteSeer}. This implies that information diffusion has different impact on publications in theoretical computer science and applied computer science. \revl{In future work we would like to probe these differences in information diffusion for various fields further. }
\begin{table}[hbt]
\begin{tabular*}{0.90\columnwidth}%
     {@{\extracolsep{\fill}}|l||lll|}
    \hline
  Venue  & Dataset & Percent. & Correlation \\
  \hline
  \hline
   Theory & ACM & 32.92\% & $-0.0709^{***}$  \\
   \& Algorithms & CiteSeer & 17.24\%  & $-0.0169^{***}$  \\
    \hline
    Distributed  & ACM & 6.39\% & $0.0223^*$ \\
    Computing & CiteSeer & 10.45\% & $-0.0018^*$  \\
    \hline
   Artificial & ACM & 5.24\% & $0.0115^*$\\
   Intelligence & CiteSeer & 8.69\% & $0.0838^{***}$ \\
   \hline
    Software & ACM & 19.54\% &  $0.0386^{***}$ \\
    Engineering & CiteSeer & 19.37\% & $0.1010^{***}$ \\
   \hline
\end{tabular*}
\caption{Correlations between community weights and normalized out-degrees of citing papers grouped by different communities.} \label{table:communitycorrelations}
\end{table}

\subsubsection{Subnetworks of papers in different time periods}

Instead of grouping papers according to their areas, we \revl{can also group them by publication date}. In order to reduce the noise introduced by the incompleteness and sparsity of the data sets, we only choose papers in the following four time periods for both the \texttt{ACM} and \texttt{CiteSeer} data sets: \revl{1980--1984, 1985--1989, 1990--1994, and 1995--1999}.

\begin{figure}[htbp]
  \includegraphics[width=0.85\columnwidth]{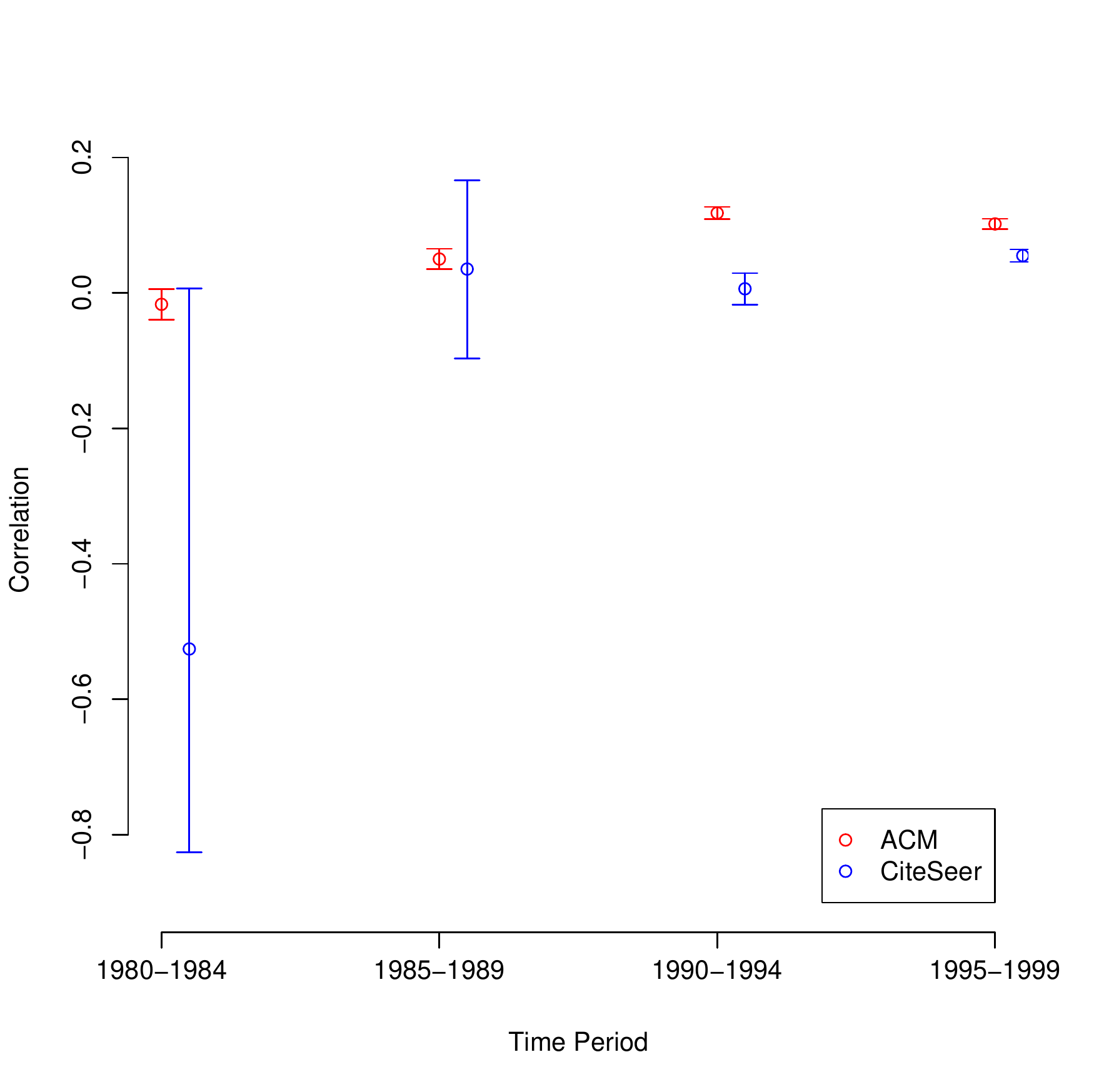} \\
  \vspace{-0.5cm}
   \caption{\revxl{Correlations between community weights and normalized out-degrees of citing papers, grouped by different time periods.}}
   \label{fig:correlationsdifftime}
   \vspace{-0.1cm}
\end{figure}

After grouping the papers according to \revl{publication date}, same as before, we select the edges with destination papers in the chosen set of papers (e.g. published between 1990 and 1994).  
We use Pearson correlations of community weights on the citation edges and the logarithm of normalized out-degrees of the destination papers, which are shown in Figure~\ref{fig:correlationsdifftime}. Although \texttt{ACM} and \texttt{CiteSeer} have different ranges of confidence intervals for the correlations, the trends of the two sets of correlations are consistent --- they are \revl{slightly} increasing as the the time periods grow more recent. \revl{Perhaps} with the research areas getting finer and deeper, it may be more difficult for researchers to keep up with, understand and cite papers in areas far from their own. At the same time, their own communities have grown and diversified to incorporate information flows from other areas, so that citing within one's area may provide adequate diversity. However, these are only speculations as to the underlying reasons why citing within one's area would be of \revl{greater} benefit to more recent papers.

\subsubsection{Subnetworks of papers versus books}

We consider one final subset of the citation graph, that of books. As we mentioned before, in \revl{the} \texttt{ACM} dataset, there are documents labeled as books or book chapters. We select these documents, and study \revl{how their citations patterns may be different from those of journal research articles or conference proceedings}. 
Since the datasets did not map books to different fields of computer science, 

we just consider the raw out-degree of books as the measure of impact and focus on the time elapsed between the publication of the book and the work it cites. We consider citations from books or book chapters to any type of publication, including papers in journals and conference proceedings.  Because books have longer reference lists (see Figure~\ref{fig:degreedistributions}(a)), any single citation is less likely to have a strong effect on the impact of the citing article. Indeed, we find that the correlation of time spans and the out-degrees of books is $-0.049^{*}$. \revxl{This is lower than the corresponding correlation between out-degrees and time spans for papers in the same dataset, which is $-0.069^{***}$.} This trend is also consistent with the fact that books are expected to cover a substantial amount of material, which may necessitate citing earlier publications. On the other hand, papers may just need to cite the most recent work they are building upon.

\section{Conclusions and future work}\label{sec:conclusion}
We analyzed a very old, regimented, and established social medium for knowledge sharing in
order to discover patterns of information flow with respect to
community structure. Consistent with prior results, we find a wide
range in the impact individual publications have. \revl{Information cascades, encompassing all chains of citations resulting from a single paper, vary dramatically in size, and only a small proportion of paper pairs are linked via cascades. In contrast the most influential papers are surprisingly interlinked.} Many publications go mostly
unnoticed, while some garner considerable attention. There are
interesting factors, relating to the citation graph, \revl{that correlate with} the popularity a given publication will enjoy. 

\revl{O}ur particular interest
is on the impact of a particular citation on the success of the
citing article. Through intensive study of two data sets of computer science publications, \texttt{ACM} and \texttt{CiteSeer}, we find that citations that occur
within communities lead to \revl{a slightly higher number of} direct
citations; and also, citing more recent 
papers corresponded to receiving more citations in turn. However, our most interesting finding is that for the most influential group of papers, this relationship was reduced or absent, allowing for the possibility that ideas across communities can lead to higher impact work. \revl{Finally, we find} that the effect of recency and community on citation structure differs among different areas of computer science and among different time periods.

In future work, we would like to expand our study to several
additional contexts, including \revxl{patent citation networks and paper citation networks of various scientific areas, in which the effect of boundary spanning information flows would be investigated.} \revl{We would also like to extend our analysis to blogs, whose} strong community structure has been observed, along with observations of information
cascades, but little is known about the effect of this community
structure on diffusion properties. \revl{However, the sparseness of citation data for blogs, and the loose relationship between them will present additional challenges}. We would also like to extend our study using textual analysis, to map specific ideas that are spreading through the citation network. In doing so, we could identify the points at which a particular idea has crossed a community boundary, and measure \revxl{whether this occasionally leads to large information cascades.}

\section*{Acknowledgements}
We would like to thank Eytan Bakshy for helpful comments and suggestions. This research was supported in part by a grant from NEC.

\bibliographystyle{aaai}
\bibliography{citations}
\end{document}